# INVESTIGATING CARGO LOSS IN LOGISTICS SYSTEMS USING LOW-COST IMPACT SENSORS

Prasang Gupta, Antoinette Young and Anand Rao

AI and Emerging Technologies, PwC, India

***ABSTRACT***

*Cargo loss/damage is a very common problem faced by almost any business with a supply chain arm, leading to major problems like revenue loss and reputation tarnishing. This problem can be solved by employing an asset and impact tracking solution. This would be more practical and effective for high-cost cargo in comparison to low-cost cargo due to the high costs associated with the sensors and overall solution. In this study, we propose a low-cost solution architecture that is scalable, user-friendly, easy to adopt and is viable for a large range of cargo and logistics systems. Taking inspiration from a real-life use case we solved for a client, we also provide insights into the architecture as well as the design decisions that make this a reality.*

***KEYWORDS***

*Asset tracking, Logistics, Cargo loss, Cargo damage, Impact sensor, Accelerometer sensor, Low-cost solution, No code AEP (Application Enablement Platform).*

## 1. INTRODUCTION

Amid the advent of globalisation, the transit of goods from one location to another is a quintessential part of any business with manufacturing or a supply chain arm. However, due to the sheer amount of logistics involved in global transportation which requires an array of different stakeholders, the handling of the actual package to be transported takes a back seat. In many cases, these packages are poorly handled or there is a hefty premium involved for ensuring proper handling of the package which makes the whole transportation process too costly for the parent business. In fact, The National Cargo Security Council (NCSC) estimates that the global financial impact of cargo loss exceeds $50 billion annually [1], moreover 50% of domestic and international insurance claims are denied [2].

The financial impact of goods damaged during transit goes beyond the cost of replacing damaged cargo [3]. There are many other factors to be considered such as interruptions to the supply chain, higher insurance claim costs, loss of productivity due to filling out claims, tracking down return orders, repackaging and shipping replacements, etc. Damaged cargo can also damage relationships among the firm, vendors and clients and can also project a negative impact on an organizations' brand value leading to further loss of sales and market share.

This serious problem of damaged goods with dire consequences for any business is surprisingly straightforward to mitigate with some care and infrastructure in place. Cargo damage can be prevented if the right measures are taken. However, prevention requires information regarding the conditions that the cargo experiences while it is in transit [4]. Increasing the visibility in the transit section of the supply chain could allow the businesses to track the amount of cargo damage occurring [5] and provide valuable information relevant to mitigating this loss. One of the possible methods is to use business frameworks and high-level risk analysis for the transit [6]. This also includes laying out operational guidance for the crew [7]. Another method is to collect the information regarding the impacts that the cargo faces while in transit as well as the geographical locations and the intensity of the impacts. This can be done by actively monitoring the cargo using sensors. We will discuss the latter in this study.

Tracking a shipment is a very common and important problem statement. There are several solutions that are present in the literature including using LoRa WAN technology [8], radio frequency identification (RFID) [9], or dynamic scheduling [10]. We explore a different strategy here and focus on capturing and tracking the impacts faced by the cargo [11] with the prime focus of keeping the solution cost effective and user friendly.

There are many types of sensors that can be used to measure impacts and vibrations which can be helpful in preventing damage to the cargo [12]. However, most of these sophisticated sensors cost upwards of a couple thousand dollars [13,14]. While this may be viable for high-value goods, this might not be cost-effective if the cost of the sensor is comparable to the cost of the cargo, which is going to be the case for most types of goods and businesses.

We discuss here a solution that we employed for one of our clients facing a very similar issue. The salient feature of the solution is that it can be scaled, it is generic enough to be extended to multiple problems and requires very low upfront capital investment, hence, is economically feasible to implement for most types of cargo. We will discuss the general solution architecture and design choices across the study, with short details and specifics about our client problem to give a flavour of a real working solution.

In the next few sections, we will first discuss the types of sensors that are viable for this kind of a problem along with the sensor that we used with a brief overview of its features. Then, we will move on to discuss the analytics behind converting the raw sensor values to meaningful information. Lastly, we will discuss how we can bring all of this together in a holistic solution architecture using an application enablement or a cloud platform.

## 2. SENSOR

We used a cost-effective accelerometer sensor to capture vibrational and/or impact data instead of sophisticated sensors in this study. This section starts off with a brief description of the accelerometer sensor. The next section details the different firmware-level changes that need to be made on the sensor to suit our specific class of problems. Finally, we will have a brief discussion on the features that were instrumental in the decision behind selecting the sensor.

### 2.1. Accelerometer Sensor

An accelerometer is a sensor which measures proper acceleration. Proper acceleration is the acceleration faced by a sensor in its own frame of reference. This implies that the acceleration recorded by the accelerometer while sitting statically would be 9.8 $m/s^2$ on the axis which is perpendicular to the ground and 0 on the other two axes. For e.g., the y axis of the accelerometer sensor in Figure 1 would register an acceleration of -9.8 $m/s^2$ while the x and the z axis would register 0 acceleration.

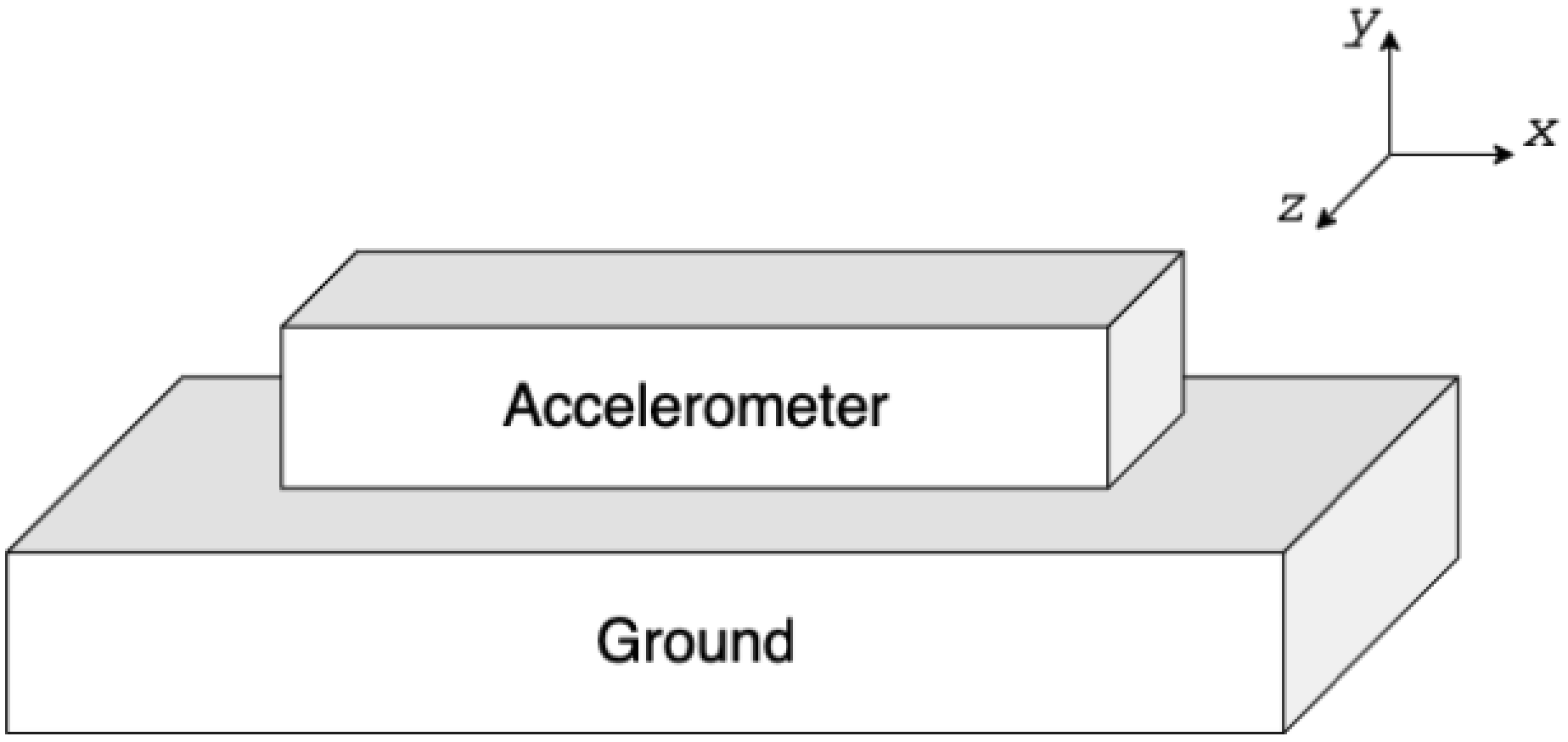

Figure 1. Accelerometer sensor placed with y axis perpendicular to the ground

The accelerometer sensors are usually configured to report the acceleration values multiple times within a second. The frequency of each sensor depends on the hardware capabilities of the sensor as well as how it is configured. Also, the maximum acceleration that can be recorded by the sensor before it is capped off, also varies between different configurations.

The accelerometer sensor that we used was housed in a multi-purpose device built by a location tracking equipment manufacturer. The device also contained a temperature sensor, a humidity sensor and GPS along with memory and communications support. This allowed the device to store collected data based on triggers and then flush it to the cloud whenever an internet connection was established. This connection establishment to the internet could be put on a schedule if needed. It also contains firmware which allows configuration of each of the sensors housed inside the device. We used this firmware to tweak the accelerometer which is discussed in the next section.

## 2.2. Firmware Changes

As explained earlier, this device collects and stores the accelerometer data multiple times within a second. The general use case does not benefit from getting instantaneous acceleration of the device, however it does by monitoring the sudden changes in acceleration which are characteristic of sudden jerks or an impact. This can be achieved by listening and recording all the values that cross a particular threshold. This threshold can be decided based on the fragility of the cargo. We optimised the threshold by performing several drop tests and calibrating the accelerometer raw values at the point where the drop was lethal to the package.

Another important aspect of the solution is to send the raw data collected by the sensors to the cloud periodically. Sending data to the cloud is a battery-intensive process, hence, it needs to be throttled to enable the device to keep running for long periods of time without a need to be recharged mid-transit. The cloud connection schedule of the device also needs to be optimised to confirm that the device remains functional throughout the transit and does not drain the battery, prompting the device to need charging. This is important from a logistics perspective as the lower the maintenance required for the device, the better.

For our case, these firmware changes allowed the device to complete the transit in its first test run. The battery life of the device can be extended from a few days to a year after changing the cloud connection frequency. This allows the implementation to work without any need of recharging of the device till the end of a cycle. Adding to that, when the demo cycle using the device in a real case study was finished, the device was still at about 30 % of battery capacity even after being used for about 4 months.

### 2.3. Sensor Requirements

As we have already discussed, there are several specific requirements for our general use case. Hence, a device needs to be selected that checks all the boxes and requirements. At the same time, it is important that the sensor remains cost-effective. Here is a summary of all the requirements that the selected sensor should be able to fulfil:

- As we are assuming that our shipping is across country borders, it can also be safely assumed that it will cover both land and sea routes. Hence, there is a requirement for the device to be able to connect to the cloud and send data while roaming on any cellular network available in a region. Connections using BLE (Bluetooth Low Energy) and WiFi won't work because of the unavailability of such networks in remote areas
- The total shipping time would be of the order of weeks. The device should last the whole journey without the need to be recharged. This would confirm hassle-free logistics and provide another layer of robustness to the solution.
- The shipping route would cover multiple nations and hence, the connectivity options should be nation-agnostic
- The data to be sent to the cloud is of the order of KBs (raw accelerometer values with timestamp and location information) and hence, the device needs to have sufficient on-board memory and should be able to support this bandwidth while sending data to the cloud

Owing to these requirements, we chose the connectivity option as LTE-m/NB-IoT with a 2G/3G fallback for the device. The device was configured to use 3G network, if available, to increase the speed of transfer. However, in the case when no 3G network was found, it would look for a 2G network and communicate through that. This network choice confirmed that:

- The device may connect to the cloud anywhere without any need for an established network (like Wi-Fi or Bluetooth)
- The bandwidth supported by 2G/3G networks is enough for sending data of the order of KBs to the cloud
- This option would also work in many regulatory domains because of the abundant presence of 2G/3G towers as opposed to only 4G networks.
- Using a 2G/3G network also enabled connectivity due to existing roaming agreements between carriers

## 3. ANALYTICS

The analytics and the transformations on the raw accelerometer data are the core of the solution of making the cost-effective sensors perform as good as sophisticated sensors for our general use case. These transformations would ideally provide us with actionable insights regarding the different impacts, including both the frequency and the level of the impact, as well as the locations of these impacts. The locations would help us in pinpointing the exact arm of transit, and impact can be mitigated by changing the transporter for that arm or by increasing investment in packaging based on the levels of impact faced.

The raw sensor data is in the form of a collection of objects. Each of these objects contains a timestamp, current battery level, accelerometer readings for each axis and additionally, readings of other sensors present in the device, if any.

When data is flushed by the device, an array of these objects are uploaded to the cloud. All of these objects have the accelerometer value above the pre-decided configured threshold. The data can be accessed for running analytics either directly through a data pull from the cloud or by putting an API (Application Programming Interface) service on top of the data storage and requesting it with configured parameters thereafter.

The device may contain other sensors apart from the accelerometer like temperature, humidity etc. We can choose to either disregard all the other sensors or spin up a dashboard around them for more insights about the transit. These sensors could be used to add further insight on the environmental factors that have a direct effect on the cargo. These can then be switched on and off in the analytics section accordingly.

The timestamp and battery information sent by the device follows a certain pattern specified by the manufacturer and would be straightforward to work upon. However, the raw accelerometer values generally come in scaled quantities. For our case, it was between -512 and 512, with 0 specifying no acceleration on that axis. The drawback of this was there was no attached physical sense to these readings and the impact values can only be used for comparison purposes.

To generate physical meaning from these values, they need to be calibrated. This was done by proposing a function that converts the raw accelerometer values into a force value in some physical units. We have chosen the 'g-force' unit for our study. The function that we used in our study was

$$V = \lambda\sqrt{x^2 + y^2 + z^2}$$

where

$V$=*Impact value in* g *force*$\lambda$=*Case specific constant* $x,y,z$=*Accelerometer raw values*

Equation 1. Calibration equation to convert raw accelerometer values to g-force units

The case specific constant mentioned in Equation 1 was calculated by performing several experiments using the sensor. Most of these experiments involved dropping the sensor from different heights and noting down the impact numbers registered by the accelerometer and then back calculating the constant value. These tests also helped in identifying the threshold g-force value that could be considered dangerous for the equipment to be tracked. This threshold would vary depending on the packaging and the fragility of the cargo.

## 4. Solution Architecture

Having all the pieces of the problem solved, this section deals with bringing them all together in a coherent architecture. There are several ways to design this. It could be an on-premises system, a cloud-based system or a hybrid between the two. Also, any platform that supports API requests and dashboarding can be used for creating the architecture. The architecture for the whole solution can be divided into 4 distinct components: hardware layer, data layer, ingestion engine, and presentation layer. This can be visualized in Figure 2. We will discuss each of these components in greater detail.

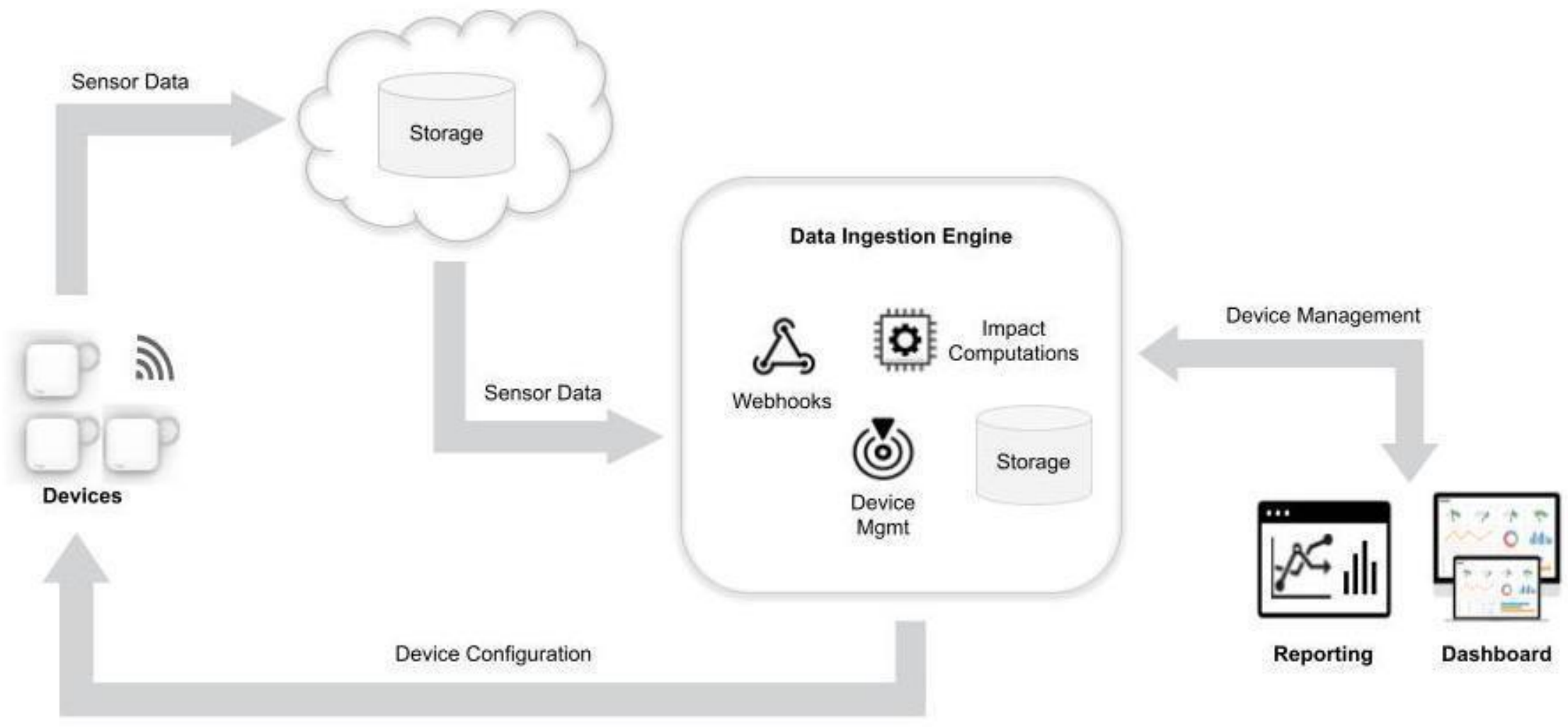

Figure 2. Overall Solution Architecture

## 4.1. Hardware Layer

The hardware layer consists of all the sensors required and the housing for the same. The purpose of this layer is to capture the values recorded by these sensors and store it in memory. Another responsibility of this layer is to connect to the cloud on a scheduled basis and flush the captured data to the cloud. This can be used either for analysis purposes or for logging and historical data collection if the analysis is done at the edge.

For our use case, we used an accelerometer housed in a pre-built key-sized device. The device also had the capability to connect with the cloud and flush its data stored in on-board memory. We configured this according to our need and that concluded our hardware layer solution.

## 4.2. Data Layer

The data layer consists of all the data that is being stored, either for archival or for analysis purposes. The responsibility of this layer is to manage all the data that is generated by the sensors and provide it on an on-demand basis.

For our use case, we configured the device to send the data to the cloud which in turn gets fed to a database. An API was built on top of that for retrieval based on timestamp range and device ID. This API was kept secure, and data was retrieved from this using proper keys for further analysis of the raw data stored.

## 4.3. Ingestion Engine

The ingestion engine consists of all components that are needed to convert the raw data into meaningful insights. This component includes compute engines, database, and ETL (Extract-Transform-Load) functions etc. The responsibility of this layer is to fetch the raw data from the data layer and make the data ready to be published by the presentation layer.

For our use case, we created an ETL engine and a database to store the processed input. The ETL engine was configured to get the data from the data layer API in short batches and update the processed database. This layer was hosted on a no-code AEP (Application Enablement Platform).

### 4.4. Presentation Layer

The presentation layer consists of all the components that the end user would directly interact with. This includes all dashboards and websites generated for the user's perusal. The responsibility of this layer is to fetch the processed data/insights from the ingestion engine and publish it in meaningful ways on a dashboard. This includes plots, buttons, colour coded depictions etc.

For our use case, we generated a dashboard and hosted it online. The dashboard generated a live time series plot of all the impacts faced by the device, which it fetched directly from the processed database. This plot was also configured to include the current general location of the device, hence enabling the user to ascertain the regions where the cargo is facing stronger impacts. The dashboard also contained other device-level information such as current battery level, last known location and the last connection timestamp with the device. A snapshot of the dashboard can be seen in Figure 3. This layer was also hosted on the same no-code AEP.

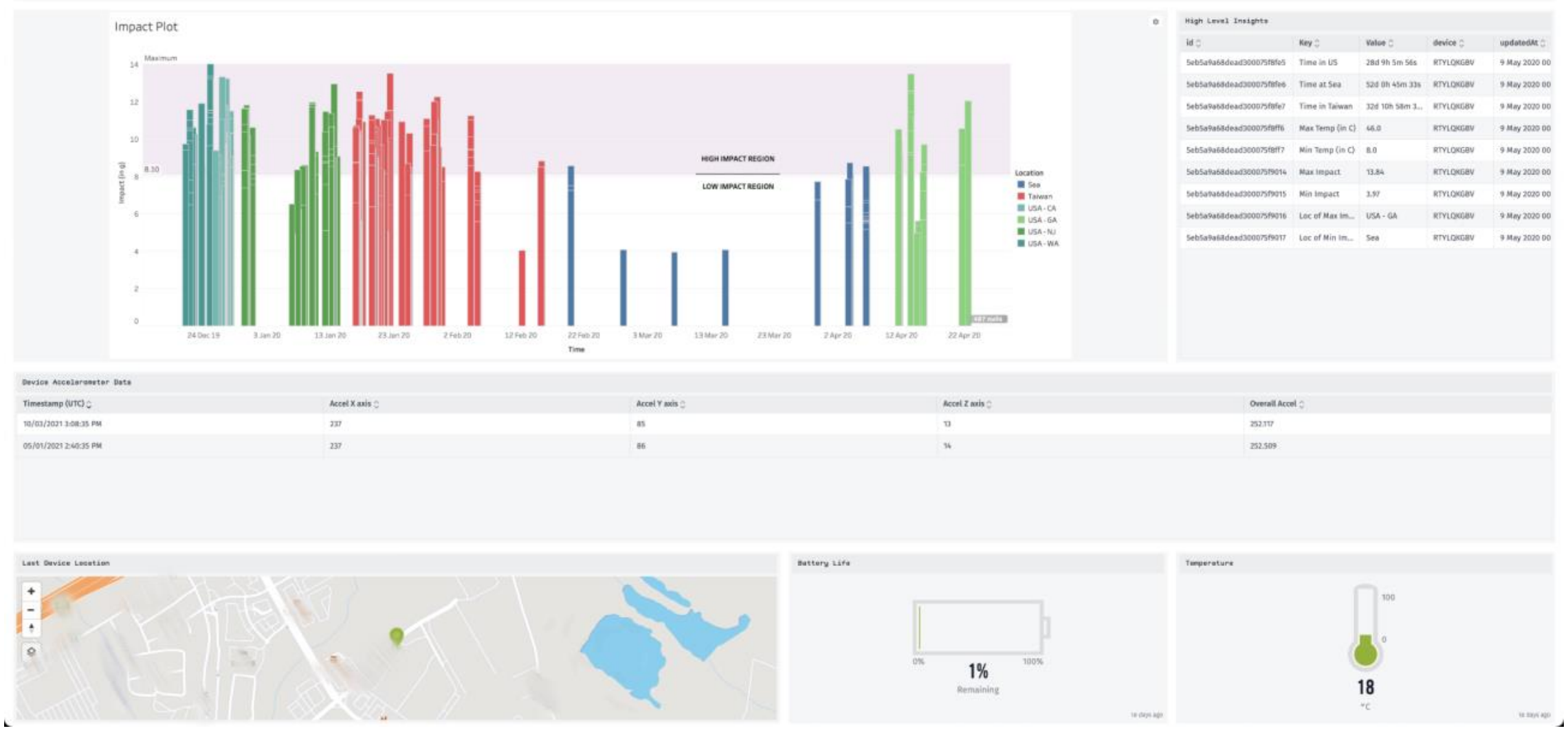

Figure 3. A snapshot of the dashboard published

## 5. Results and Discussion

In the previous sections, we have mentioned and discussed in detail a method that can be employed for easy asset and impact tracking. We have also covered the key design aspects involved along with general guidance for making these decisions and a brief overview of the pros and cons of each, with respect to our particular use case. We will continue this discussion here and highlight the salient points of the solution developed and how it was helpful for the client.

The live dashboard generated for the client, as seen in Figure 3 was hosted online giving the client easy access for tracking all their current shipments in the same place. Apart from all the fields provided regarding the general information of the device, specialised impact plots were created highlighting the impact faced by the device in different legs of its journey along with the demarcation of high/low impact. This was used to ascertain the regions where it was most probable for the cargo to get damaged. This plot can be seen in Figure 4.

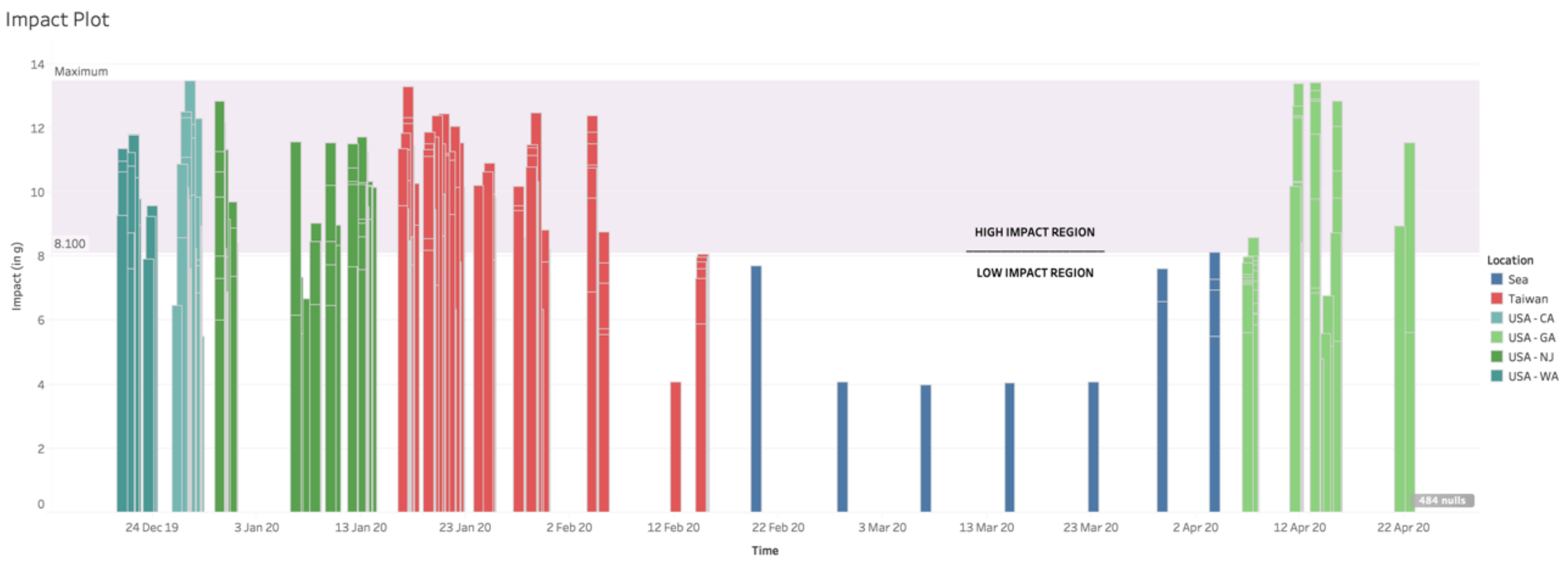

Figure 4. A snapshot of the impact plot generated

The plot generated in Figure 4 gave invaluable insights into the journey of the shipment from the manufacturer to the warehouse. It was concluded from the charts that most of the high impacts faced by the device were on land routes (red bars on the left are the impact readings when the shipment was shipped from the manufacturer and green bars on the right are the impact readings when the cargo was transported from the docks to the warehouse) rather than on sea routes (blue bars represent cargo travelling on container ship). This provided transparency to the client on their whole supply chain, creating value by assisting them in making decisions to reduce cargo damage.

Overall, this solution was built keeping the ease of logistics in mind, which included one-time use devices with no hassle of recovering them from cargo, a device that would not require charging within the transit period, and ease of placing the device on the cargo with no special instructions/handling directions etc. This proved to be very useful in terms of the viability and ease of adoption and scalability of the solution.

There are certain limitations to this type of architecture in the logistics space. Most importantly, this creates issues when there is a need for recovery of the tracking devices as this is not covered. Also, tuning frequency of the data collection and other device parameters needs some prior experience in the IoT domain to get the optimal values as recreating real-life situations is not possible. Another limitation of this method is the long feedback cycle as the parameters can only be optimised after a single shipment is complete. Hence, getting the desired parameters could take 2-3 shipment cycles, assuming that the cycle remains the same every time.

## 6. Conclusion

Damaged or lost cargo can have a negative impact on every business with a supply chain arm. This impact can be both short term like product loss, replacement inventory, time lost in filing claims etc. and long term like loss of clients or market share, loss of trust in the brand etc. As discussed, a major share of damage or loss of cargo happens during its transportation.

Reducing in-transit damage by leveraging the latest technologies for monitoring impacts and vibrations can offer several benefits to the business and help them in the long run. Some of these benefits are:

- Real-time tracking of impact, vibrations etc. to the cargo increases visibility through the transit route and helps determine where and when damage is occurring even when out of coverage.
- Monitoring provides data that can be used to help prevent damage in the future by using proper packaging for the cargo.
- Understanding the trends in damage occurrence, delays, and other issues can help in improving performance leading to better customer service and satisfaction

This work, with careful calibration and design choices, can further be expanded from impact tracking to much more sophisticated vibration tracking. This would open up arrays of opportunity for this system to be implemented. Also, other sensors can be used in conjunction with the accelerometer sensor to solve multiple complex problems.

## AUTHORS

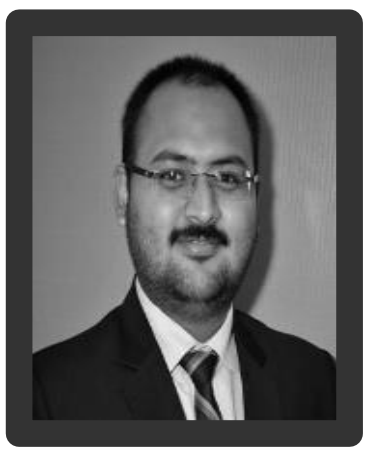

**Prasang Gupta** is an Experienced Associate in PwC's AI and Emerging Technologies team. He has 2 years of experience in solving complex real-world problems with cutting edge ML and AI inspired innovative solutions. He earned his Master of Technology and Bachelor of Technology in Chemical Engineering from Indian Institute of Technology Kanpur.

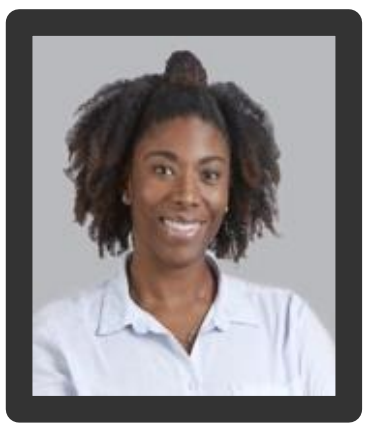

**Antoinette Young** is a manager in PwC's AI and Emerging Technologies team. In this role she does research and develops prototypes focused on Internet of Things (IoT) technologies. Antoinette is passionate about all things robotics. She earned a Master of Science in Information Technology from Nova Southeastern University and a Bachelor of Science in Computer Science from Florida Atlantic University.

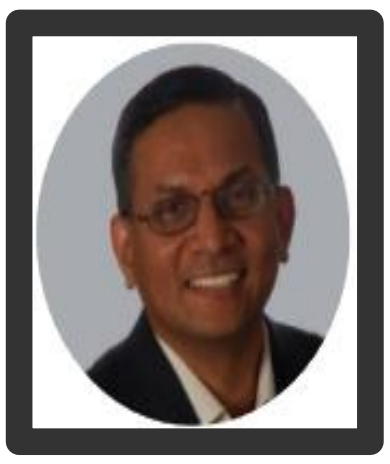

**Anand Rao** is a principal in PwC's AI and Emerging Technologies team and the global AI lead with over 30 years of industry and consulting experience, helping senior executives structure, solve and manage critical issues facing their organizations. He holds a MSc (Tech) in Computer Science from Birla Institute of Technology and Science in India and PhD in Artificial Intelligence from University of Sydney, where he was awarded the University Postgraduate Research Award. also awarded an MBA from Melbourne Business School with Distinction.

*PwC refers to the US member firm, and may sometimes refer to the PwC network. Each member firm is a separate legal entity. Please see www.pwc.com/structure for further details. This content is for general information purposes only, and should not be used as a substitute for consultation with professional advisors.*